\documentclass[fleqn,10pt,twocolumn]{wlscirep}

\usepackage{amsmath,amssymb}
\usepackage{bm}
\usepackage{xcolor}
\usepackage{subfigure}
\usepackage{ragged2e}
\usepackage{authblk}
\usepackage{flushend} 
\usepackage{tikz}
\usetikzlibrary{patterns}
\usetikzlibrary{calc,arrows,decorations.markings}

\newcommand{\ie}{{\it i.e.~}}
\newcommand{\eg}{{\it e.g.~}}

\newcommand{\etcn}{{\it etc\dots}}
\newcommand{\vs}{{\it vs.~}}
\newcommand{\vv}{{\it vice versa}}

\newcommand{\Slabel}[1]{\label{sec:#1}} 

\newcommand{\Sref}[1]{``\textbf{\nameref{sec:#1}}''}

\newcommand{\elabel}[1]{\label{eq:#1}}

\newcommand{\Eref}[1]{Eq.~(\ref{eq:#1})}

\newcommand{\flabel}[1]{\label{fig:#1}}

\newcommand{\Fref}[1]{Fig.~\ref{fig:#1}}
\newcommand{\BFref}[1]{Figure~\ref{fig:#1}}

\newcommand{\ave}[1]{\left\langle #1 \right\rangle}
\newcommand{\abs}[1]{{\left|#1\right|}}
\newcommand{\plaind}{\mathrm{d}}
\newcommand{\dint}[1]{\mathchoice{\!\plaind#1\,}{\!\plaind#1\,}{\!\plaind#1\,}
{\!\plaind#1\,}}

\title{Spatial Patterns in Urban Systems}

\author[1,2,C]{\textsc{Hoai Nguyen HUYNH}\thanks{\texttt{https://%
sites.google.com/site/nelive/}}}
\author[3]{\textsc{Evgeny MAKAROV}\thanks{\texttt{http://www.baseride.com/}}}
\author[1]{\textsc{Erika Fille LEGARA}\thanks{\texttt{http://www.erikalegara.%
net/}}}
\author[1,2]{\textsc{Christopher MONTEROLA}\thanks{\texttt{http://www.%
chrismonterola.net/}}}
\author[2,4]{\textsc{Lock Yue CHEW}\thanks{\texttt{http://www3.ntu.edu.sg/home/%
lockyue/}}}

\affil[1]{Institute of High Performance Computing, Agency for Science Technology
and Research, Singapore}
\affil[2]{Complexity Institute, Nanyang Technological University, Singapore}
\affil[3]{Baseride Technologies, Singapore}
\affil[4]{School of Physical and Mathematical Sciences, Nanyang Technological
University, Singapore}

\affil[C]{huynhhn@ihpc.a-star.edu.sg}


\begin{abstract}
Understanding the morphology of an urban system is an important step toward
unveiling the dynamical processes of its growth and development. At the
foundation of every urban system, transportation system is undeniably a crucial
component in powering the life of the entire urban system. In this work, we
study the spatial pattern of $73$ cities across the globe by analysing the
distribution of public transport points within the cities. The analysis reveals
that different spatial distributions of points could be classified into four
groups with distinct features, indicating whether the points are clustered,
dispersed or regularly distributed. From visual inspection, we observe that the
cities with regularly distributed patterns do not have apparent centre in
contrast to the other two types in which star-node structure, \ie monocentric,
can be clearly observed. Furthermore, the results provide evidence for the
existence of two different types of urban system: well-planned and organically
grown. We also study the spatial distribution of another important urban entity,
the amenities, and find that it possesses universal properties regardless of the
city's spatial pattern type. This result has one important implication that at
small scale of locality, the urban dynamics cannot be controlled even though the
regulation can be done at large scale of the entire urban system. The relation
between the distribution of amenities within the city and its spatial pattern is
also discussed.
\end{abstract}

\begin{document}

\flushbottom
\maketitle

\thispagestyle{empty}

\section{Introduction}

Study of urban systems---how they form and develop---constitutes an important
portion of human knowledge, not only because it is about our own physical space
of daily living but also for understanding the underlying mechanisms of human
settlement and civilisation on the Earth's surface that may be fundamentally
similar to other forms of organisation like biological cells in our body or
animal colonies. Urban systems, or ``cities'' in modern terms, are typical
example of highly complex systems \cite{2005@Batty,2007@Bettencourt.etal,
2010@Bettencourt.etal,2010@Bettencourt.West,2013@Bettencourt,2013@Batty_a,
2013@Batty_b} in which overwhelmingly many agents are interacting in non-trivial
and non-linear manners over a wide spectrum of spatial and temporal scales. The
results of such tangled interactions are the emergence of unexpected global
patterns that cannot be solely derived from the local knowledge of individual
agents. Among these complex patterns are the spatial patterns delineated by
the physical locations and shapes of urban entities like buildings, parks,
lakes or infrastructure \etcn, \ie the urban morphology
\cite{1994@Batty.Longley}. A good understanding of the morphology of an urban
system provides us with the comprehension of its current status of development
or even the living condition of people inside it. For example, the number of
residential buildings is a good gauge of the population size and the population
density measures the crowdness that every resident has to experience in his or
her daily life, or the infrastructure is an indicator of how well the city is
doing in terms of economy.

In the recent years, with the availability of technologies and new mapping
techniques, various forms of data on urban systems have been collected
\cite{OSM}. These data sets have enabled researchers to gain a deeper insight
into the spatial structures existing in urban systems. One such data set is on
the street and road networks which form the backbone of any urban system, and
therefore, contain rich information about how the city is organised. A good
amount of research have been performed on understanding the pattern of streets
in many different cities around the world \cite{2007@Bin,
2008@Barthelemy.Flammini,2012@Strano.etal,2013@Barthelemy.etal,
2013@Gudmundsson.etal,2013@Strano.etal,2014@Louf.Barthelemy}. However, there are
shortcomings of studying urban morphology based on street network when it has
been noticed that the streets are not always well defined \cite{2006@Porta.etal,
2014@Louf.Barthelemy}. Public transport points, conversely, are well defined and
can gauge the level of socio-technical development in an urban system as they
represent the degree of mobility activities taking place within the urban
system. Furthermore, public transport network is design to serve the residents
of the city to perform activies on every aspect of daily life (going to school,
commuting to work, shopping, entertainmment \etcn), and therefore, can be used
as a good proxy of the residential distribution within a city. At the interplay
of these factors, spatial pattern of distribution of transport points can
provide us with rich insights into the morphology or, in some cases, even the
morphogenesis of an urban system.

In this study, we will explore the spatial patterns encompassed in urban systems
by analysing the pattern of spatial distribution of transport points (bus stops)
in their public transport network of $73$ cities around the world (see
\Sref{data}). The analysis shall reveal that there exists a typical value of
distance among transport points within each city. This characteristic distance
reflects the accessibility of each transport point and connectivity of them as
an entire network, and hence, can be employed in measuring the physical area of
the transport network's coverage. Interestingly, this area is shown to exhibit a
scaling relation against the characteristic distance with a non-trivial value of
the scaling exponent. Furthermore, the spatial distribution pattern of the
transport points can be quantified and shown to belong to two main groups in
which the points are either approximately equidistant or they are distributed
apart with multiple length scales. The first group contains cities that appear
to be well-planned, \ie organised type, while the second consists of cities that
tend to spread themselves over a large area and possess non-uniform spatial
density of urban entities at different length scales, \ie organic type. In
addition to public transport network, we also look at the distribution of
amenities within each city to investigate the relation between these two types
of urban entity. We first find that the distance between amenities to their
nearest transport points within a city follows a robust exponential distribution
across all the cities considered, regardless of the city's type being organised
or organic. Subsequently, we observe a clear quantifiable relation between this
amenity-transport point distance and the density of transport points; and the
type of the city can also be seen in this relation.

\section{Results}

Using a method of cluster analysis inspired by percolation theory
\cite{1994@Stauffer.Aharony}, we are able to characterise the spatial pattern of
public transport points in urban systems. The spatial spattern is characterised
by quantifying the size and area of a dominant cluster of transport points as
functions of a distance parameter. This distance parameter (also the buffer
radius) $\rho$ represents the extent of vicinity around every transport point in
the system, hence, the point's connectivity. Larger value of $\rho$ means the
neighbourhood of a point is extended, and therefore, can encompass more points
within it. A pair of points are said to belong to the same cluster if and only
if the Euclidean distance between them is less than or equal to $\rho$. The size
of a cluster is defined as the number of points in the cluster while its area
the union of area of circles of uniform radius $\rho$ centred at the points in
the cluster. A dominant cluster is the cluster with either largest size
$\xi_{max}(\rho)$ or largest area $A_{max}(\rho)$. The clusters in the two
occasions are not necessarily the same one. For simplicity of all discussions
below, unless stated explicitly, descriptions for cluster size $\xi$ also hold
for cluster area $A$.

When $\rho$ is small, the number of clusters $\eta(\rho)$ is large because most
of the points are not connected and they form their own clusters. As $\rho$
increases, $\eta(\rho)$ decreases monotonically  because of merger of small
clusters. In fact, it is a step function because the pairwise distances between
points are discrete in value. On the other hand, the size of the largest cluster
$\xi_{max}(\rho)$ increases monotonically as $\rho$ increases. Again, it is also
a step function, but we assume in this study that the profiles $\xi_{max}(\rho)$
and $A_{max}(\rho)$ can be approximated by well behaved and smooth functions so
that their derivatives exist at all points (see \Sref{characteistic_distances}
and \Sref{spatial_patterns}). In the regime of small buffer radius $\rho$,
$\xi_{max}(\rho)$ slowly increases because the clusters are still disjoint. As
$\rho$ enters an intermediate regime, $\xi_{max}(\rho)$ increases faster than it
does in the small-$\rho$ regime. This is when the larger clusters merge together
making the significant expansion in size of the largest cluster. As $\rho$
increases further, there is no further significant change to the size of the
largest cluster as it has encompassed most of the points in the domains, leaving
only minor portions surrounding. The intermediate regime of $\rho$, therefore,
could be seen as a region of ``phase transition'', similar to that in physics
\cite{1972--2001@Domb.etal}, particularly percolation
\cite{1994@Stauffer.Aharony}.

In percolation, every point is assigned a variable called the percolating
probability that controls the ability of one site to connect to another in the
domain. The higher the percolating probability is, the easier the site is
connected to others, and \vv. The percolating probability is then viewed as a
control parameter in the system. The transition occurs when the control
parameter is adjusted to a critical value (called the critical point) at which
the system transits from one state (or phase) to another with distinct
properties, namely the non-percolating and percolating phases respectively at
low and high percolating probabilities. The behaviours of the system approaching
the critical point can then be used to classify the system, \ie identifying its
universality class \cite{1999@Stanley}.

Applying this idea to our system of transport points within a city, the buffer
radius $\rho$ could be viewed as the control parameter. For small values of
$\rho$, the system is in \emph{segregate} phase, while it is in \emph{aggregate}
phase for large values of $\rho$. The transition from one phase to the other
takes place in the intermediate regime of $\rho$. The manner in which the
profiles of the largest system size $\xi_{max}(\rho)$ and area $A_{max}(\rho)$
transit through this region can characterise how the transport points are
distributed within a city.

\subsection{Characteristic distances among transport points}
\Slabel{characteistic_distances}

The intermediate regime of $\rho$ can be identified and characterised by
analysing the first derivative $\displaystyle\xi'_{max}(\rho)=
\frac{\dint\xi_{max}(\rho)}{\dint\rho}$ of the cluster size. This quantity,
which is interpreted as the rate of cluster size growth per unit of buffer
radius, produces a peak every time a jump occurs in the cluster size
$\xi_{max}(\rho)$, \ie when the cluster merges with others and grows. Every
peak in $\xi'_{max}(\rho)$, therefore, signifies the existence of one or a few
clusters of points located at a farther distance beyond those in the current
largest cluster that is being traced. As a result, this would provide us with
the information on the length scales of distribution of points within the set.
It can be easily seen that if there are many peaks, the points are distributed
in clusters that are apart with different distances; whereas the existence of
few peaks implies a uniform distribution of points that are (approximately)
equidistant from one another. In either case, it is without doubt that there
exists a characteristic distance in the spatial distribution of points. This
characteristic distance should tell us the length scale above which the points
are (largely) connected and below which they are disconnected.

Since all peaks in the derivative $\xi'_{max}(\rho)$ contribute to the growth of
the cluster $\xi_{max}(\rho)$, a measure of the characteristic distance
$\rho_\xi^\star$ must take into account the effects of all of them. However, a
high peak indicates a more significant increase in cluster size (a major merger)
than that indicated by a lower one. Hence, the average of all values of
$\rho_{\xi,i}^\dagger$ at which a peak $i$ occurs, weighted by the height
$\displaystyle\xi'_{max}(\rho_{\xi,i}^\dagger)=
\left.\frac{\dint\xi_{max}(\rho)}{\dint\rho}\right|_{\rho=\rho_{\xi,i}^\dagger}$
of the peaks, is an appropriate measure of this characteristic distance, \ie
\begin{equation}
\rho_\xi^\star =
\frac{\sum_i{\xi'_{max}(\rho_{\xi,i}^\dagger)\rho_{\xi,i}^\dagger}}
{\sum_i{\xi'_{max}(\rho_{\xi,i}^\dagger)}}\text{.}
\end{equation}
Similarly, we have for the cluster area
\begin{equation}
\rho_A^\star =
\frac{\sum_i{A'_{max}(\rho_{A,i}^\dagger)\rho_{A,i}^\dagger}}
{\sum_i{A'_{max}(\rho_{A,i}^\dagger)}}\text{.}
\end{equation}
The analyses of peaks in size profile $\xi'_{max}(\rho)$ and area profile
$A'_{max}(\rho)$ provide different perspectives on the spatial distribution of
points. The size quantifies the number of points with respect to the distance
while the area further takes into account the relative position of the points.
The two are not redundant but rather, one is complementary to the other. This
comes to light in the next section \Sref{spatial_patterns} when the combination
of the two allows us to classify distinct types of distribution of points.

It is noteworthy that $\rho_\xi^\star$ and $\rho_A^\star$ are different from the
average of pairwise distance among all points in the set because they encode the
connectivity of the points in terms of spatial distribution. In other words, the
two characteristic distances are measure of typical distance between points in
the set in the perspective of global connectivity of all points. In the context
of transport points, they translate to the distance one has to traverse to get
from one point to another in order to explore the entire system. It then follows
that a large value of characteristic distance implies a sparsely distributed set
of points which could reflect a poorly covering network of transport. This
agrees with a low density of points per unit area (see
\Sref{density_distance_relation} for more details). For the $73$ cities
considered in this study, $\rho_\xi^\star$ is found to be in the range
$200-1200$m with most cities having $\rho_\xi^\star$ in the range $200-500$m.
The ranges are $200-1700$m and $300-500$m for $\rho_A^\star$.

\subsection{Spatial patterns in urban systems}
\Slabel{spatial_patterns}

The characteristic distances introduced above tell us whereabout the transitions
of cluster size and area take place but they do not tell us how the size and
area of the cluster transit from small to large value, \ie how the cluster
grows. This, however, can be easily characterised by further exploiting the
analysis of peaks in $\xi'_{max}(\rho)$ (and $A'_{max}(\rho)$). It is a matter
of fact that if the cluster grows rapidly through the transition, there are very
few peaks in $\xi'_{max}(\rho)$, all of which are sharp and localised. On the
other hand, the peaks are scattered over a wide range of $\rho$ should the
cluster gradually grow. The standard deviation of the location
$\rho_{\xi,i}^\dagger$ of the peaks, or the spread of transition $\sigma_\xi$
for cluster size, is a good measure of such scattering. However, a low peak that
is distant from a group of localised high peaks should not significantly enlarge
the spread. Therefore, the standard deviation of $\rho_{\xi,i}^\dagger$ needs to
be weighted by the height $\xi'_{max}(\rho_{\xi,i}^\dagger)$ of the peaks, \ie
\begin{equation}
\elabel{characteristic_distance}
\sigma_\xi = \sqrt{\frac{\sum_i{\xi'_{max}(\rho_{\xi,i}^\dagger)
\left(\rho_{\xi,i}^\dagger-\rho_\xi^\star\right)^2}}
{\sum_i{\xi'_{max}(\rho_{\xi,i}^\dagger)}}}\text{.}
\end{equation}
Similarly, we have the spread of transition for cluster area
\begin{equation}
\elabel{spread_of_transition}
\sigma_A = \sqrt{\frac{\sum_i{A'_{max}(\rho_{A,i}^\dagger)
\left(\rho_{A,i}^\dagger-\rho_A^\star\right)^2}}
{\sum_i{A'_{max}(\rho_{A,i}^\dagger)}}}\text{.}
\end{equation}

The combination of these two spreads of transition enables us to characterise
the pattern of different types of spatial point distribution by interpreting
different regions of the $(\sigma_\xi,\sigma_A)$ diagram (see
\Fref{sssa_diagram}). There are four different types of distribution that can be
identified using the spreads of transition in size $\sigma_\xi$ and area
$\sigma_A$. The first one is the region of small $\sigma_\xi\approx\sigma_A$ in
which both $\xi_{max}(\rho)$ and $A_{max}(\rho)$ exhibit a sharp rise. The
second one is the stripe of medium-to-large $\sigma_\xi\approx\sigma_A$ in which
$\xi_{max}(\rho)$ and $A_{max}(\rho)$ exhibit gradual increase and almost every
peak in $\xi'_{max}(\rho)$ has a respective peak in $A'_{max}(\rho)$. The third
one is the region of $\sigma_\xi\gg\sigma_A$ in which the peaks in
$\xi'_{max}(\rho)$ tend to spread over a wider range of $\rho$ than those in
$A'_{max}(\rho)$. The last one is the region of $\sigma_\xi\ll\sigma_A$ in which
the peaks in $\xi'_{max}(\rho)$ tend to be more localised than those in
$A'_{max}(\rho)$.

\begin{figure*}[ht!]
\centering
\begin{tikzpicture}
\draw[->,line width=2,>=latex] (0,0) -- (8,0);
\draw[->,line width=2,>=latex] (0,0) -- (0,8);
\draw (4,-1) node {\LARGE$\sigma_\xi$};
\draw (-1,4) node [rotate=90] {\LARGE$\sigma_A$};

\draw[dashed,very thick] (1,1) circle (0.5cm);
\draw[->,line width=1,>=latex] (-1,1) -- (0.8,1);
\draw (-2.5,1) node {\begin{tabular}{c}regularly distributed\\at single length
scale\\small $\sigma_\xi\approx\sigma_A$\end{tabular}};

\draw[dashed,very thick,rotate=45] (5,0) ellipse (3cm and 0.5cm);
\draw[->,line width=1,>=latex] (6.5,4.5) -- (4.7,4.5);
\draw (8.3,4.5) node {\begin{tabular}{c}regularly distributed\\at multiple
length scales\\medium-to-large $\sigma_\xi\approx\sigma_A$\end{tabular}};

\draw[dashed,very thick] (6,2.2) ellipse (2cm and 1.5cm);
\draw (6,2.2) node {\begin{tabular}{c}clustered\\pattern\\
$\sigma_\xi\gg\sigma_A$
\end{tabular}};
\draw[->,line width=1,>=latex] (3,2) -- (4,1);
\draw (3.2,0.8) node {more clustered};

\draw[dashed,very thick] (2.2,6) ellipse (2cm and 1.5cm);
\draw (2.2,6) node {\begin{tabular}{c}dispersed\\pattern\\
$\sigma_\xi\ll\sigma_A$
\end{tabular}};
\draw[->,line width=1,>=latex] (2,3) -- (1,4);
\draw (1.8,4.2) node {more dispersed};

\draw[dashed] (0,0) -- (7,7);
\draw (7,7) node {$\sigma_\xi=\sigma_A$};
\end{tikzpicture}
\caption{\flabel{sssa_diagram}Interpretation of different patterns of spatial
point distributions given different values of the pair $(\sigma_\xi,\sigma_A)$.}
\end{figure*}
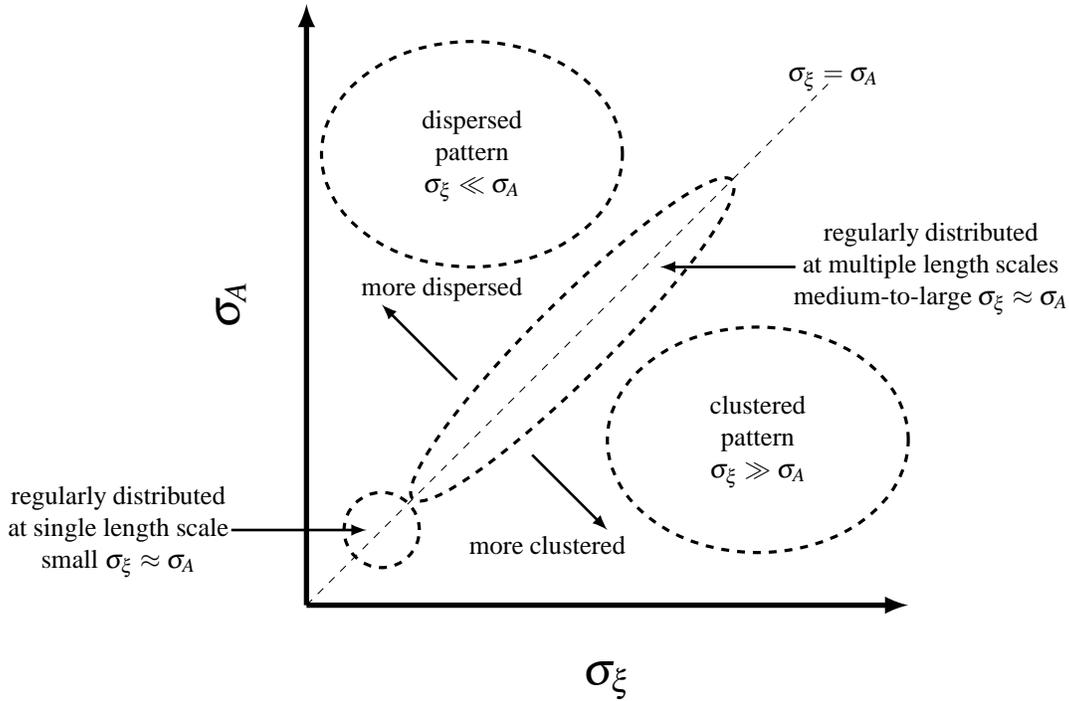

\begin{figure*}[ht!]
\centering
\includegraphics[width=1\textwidth]{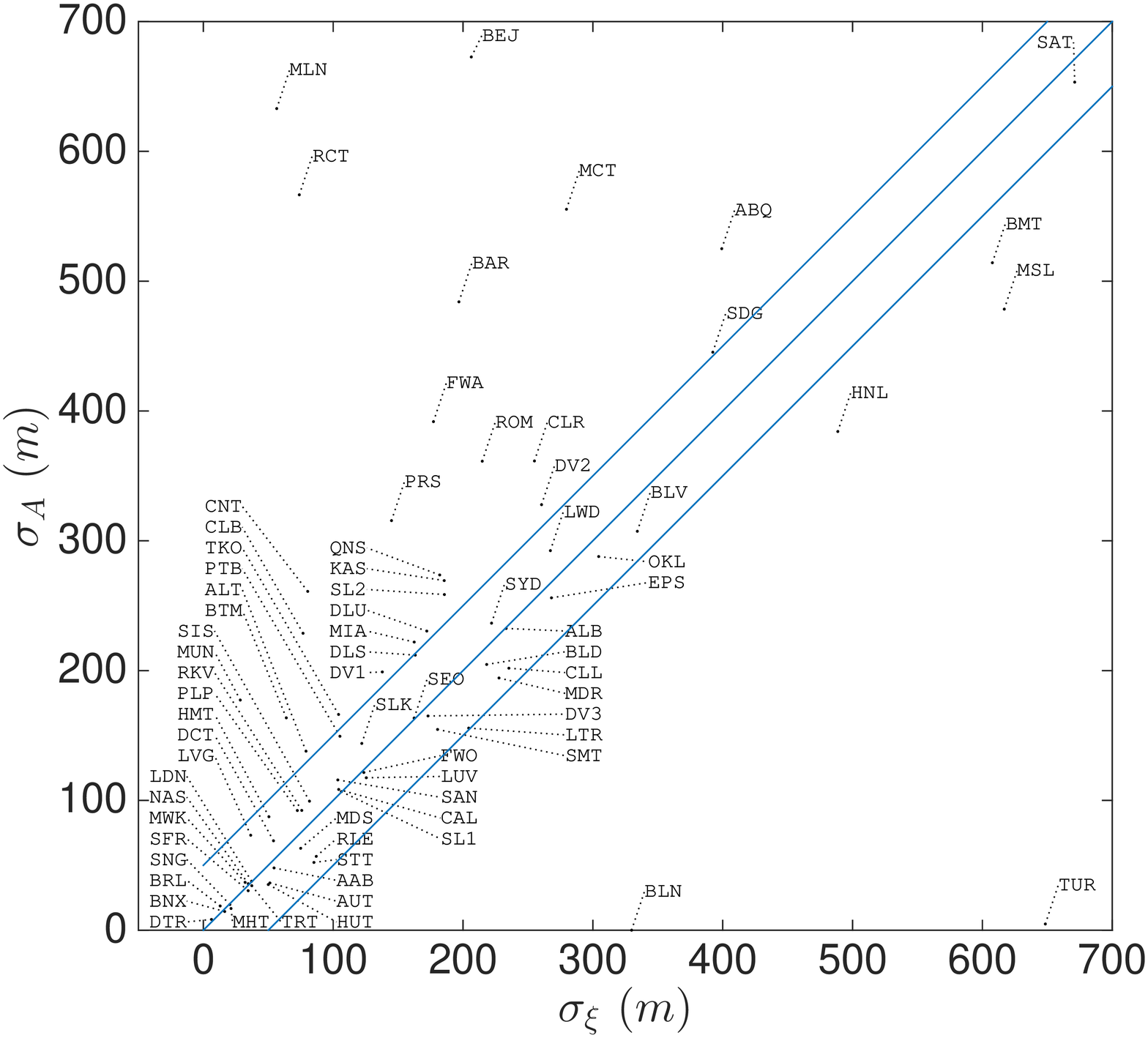}
\resizebox{\textwidth}{!}{%
\begin{tikzpicture}
\draw (0,0) node  {\begin{tabular}{llllll}
\texttt{AAB:} Ann Arbor (US) & \texttt{BTM:} Baltimore (US) & \texttt{EPS:}
Epsom (NZ) & \texttt{MDR:} Madrid (ES) & \texttt{QNS:} Queens (US) &
\texttt{SLK:} Salt Lake City (US) \\
\texttt{ABQ:} Albuquerque (US) & \texttt{CAL:} Calgary (CA) & \texttt{FWA:} Fort
Wayne (US) & \texttt{MDS:} Madison (US) & \texttt{RCT:} Rochester (US) &
\texttt{SMT:} Sacramento (US) \\
\texttt{ALB:} Albany (US) & \texttt{CLB:} Columbus (US) & \texttt{FWO:} Fort
Worth (US) & \texttt{MHT:} Manhattan (US) & \texttt{RKV:} Rockville (US) &
\texttt{SNG:} Singapore (SG) \\
\texttt{ALT:} Atlanta (US) & \texttt{CLL:} Cleveland (US) & \texttt{HMT:}
Hamilton (CA) & \texttt{MIA:} Miami (US) & \texttt{RLE:} Raleigh (US) &
\texttt{STT:} Stockton (US) \\
\texttt{AUT:} Austin (US) & \texttt{CLR:} Colorado Springs (US) & \texttt{HNL:}
Honolulu (US) & \texttt{MLN:} Milan (IT) & \texttt{ROM:} Rome (IT) &
\texttt{SYD:} Sydney (AU) \\
\texttt{BAR:} Barcelona (ES) & \texttt{CNT:} Cincinnati (US) & \texttt{HUT:}
Houston (US) & \texttt{MSL:} Marseille (FR) & \texttt{SAN:} Santa Ana (US) &
\texttt{TKO:} Tokyo (JP) \\
\texttt{BEJ:} Beijing (CN) & \texttt{DCT:} Decatur (US) & \texttt{KAS:} Kansas
City (US) & \texttt{MUN:} Munich (DE) & \texttt{SAT:} San Antonio (US) &
\texttt{TRT:} Toronto (CA) \\
\texttt{BLD:} Boulder (US) & \texttt{DLS:} Dallas (US) & \texttt{LDN:} London
(UK) & \texttt{MWK:} Milwaukee (US) & \texttt{SDG:} San Diego (US) &
\texttt{TUR:} Turin (IT) \\
\texttt{BLN:} Berlin (DE) & \texttt{DLU:} Duluth (US) & \texttt{LTR:} Little
Rock (US) & \texttt{NAS:} Nassau County (US) & \texttt{SEO:} Seoul (KR) & \\
\texttt{BLV:} Belleville (US) & \texttt{DTR:} Detroit (US) & \texttt{LUV:}
Louisville (US) & \texttt{OKL:} Oakland (US) & \texttt{SFR:} San Francisco (US)
& \\
\texttt{BMT:} Bremerton (US) & \texttt{DV1:} Denver 1 (US) & \texttt{LVG:} Las
Vegas (US) & \texttt{PLP:} Pinellas Park (US) & \texttt{SIS:} Staten Island (US)
& \\
\texttt{BNX:} Bronx (US) & \texttt{DV2:} Denver 2 (US) & \texttt{LWD:} Lakewood
(US) & \texttt{PRS:} Paris (FR) & \texttt{SL1:} St Louis 1 (US) & \\
\texttt{BRL:} Brooklyn (US) & \texttt{DV3:} Denver 3 (US) & \texttt{MCT:}
Manchester (UK) & \texttt{PTB:} Pittsburg (US) & \texttt{SL2:} St Louis 2 (US) &
\\
\end{tabular}};
\end{tikzpicture}}
\caption{\flabel{ss_sa}Types of spatial distribution of transport points in
cities across the globe. The three reference lines are $\sigma_A=\sigma_\xi$,
$\sigma_A=\sigma_\xi+50$ and $\sigma_A=\sigma_\xi-50$.}
\end{figure*}

In this analysis, for practical purpose, the regions are determined by a
$50$-meter rule. According to that rule, $\sigma_\xi,\sigma_A<50$ constitute the
small $\sigma_\xi\approx\sigma_A$ region, $\abs{\sigma_\xi-\sigma_A}<50$
($\sigma_\xi,\sigma_A>50$) constitute the medium-to-large
$\sigma_\xi\approx\sigma_A$ region, $\sigma_\xi-\sigma_A>50$ constitute the
$\sigma_\xi\gg\sigma_A$ region and $\sigma_\xi-\sigma_A<50$ constitute the
$\sigma_\xi\ll\sigma_A$ region.

\begin{figure*}[ht!]
\centering
\subfigure[\flabel{Brooklyn}Brooklyn, New York, USA. The spreads of transition
for the size $\sigma_\xi$ and area $\sigma_A$ are both small. This is an example
of single-scale regularly distributed pattern, \ie grid.]{%
\includegraphics[width=0.355\textwidth]{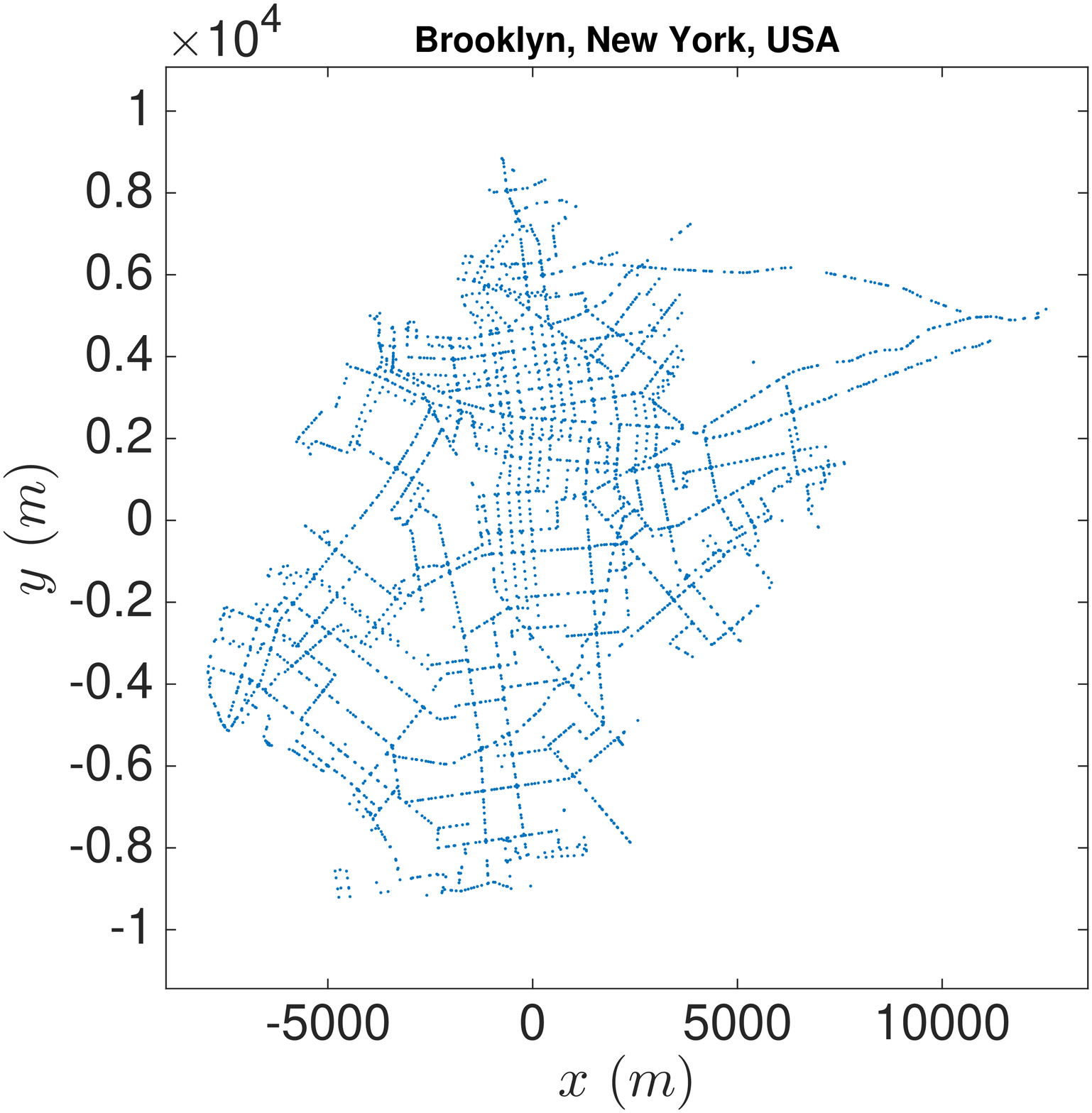}
\includegraphics[width=0.645\textwidth]{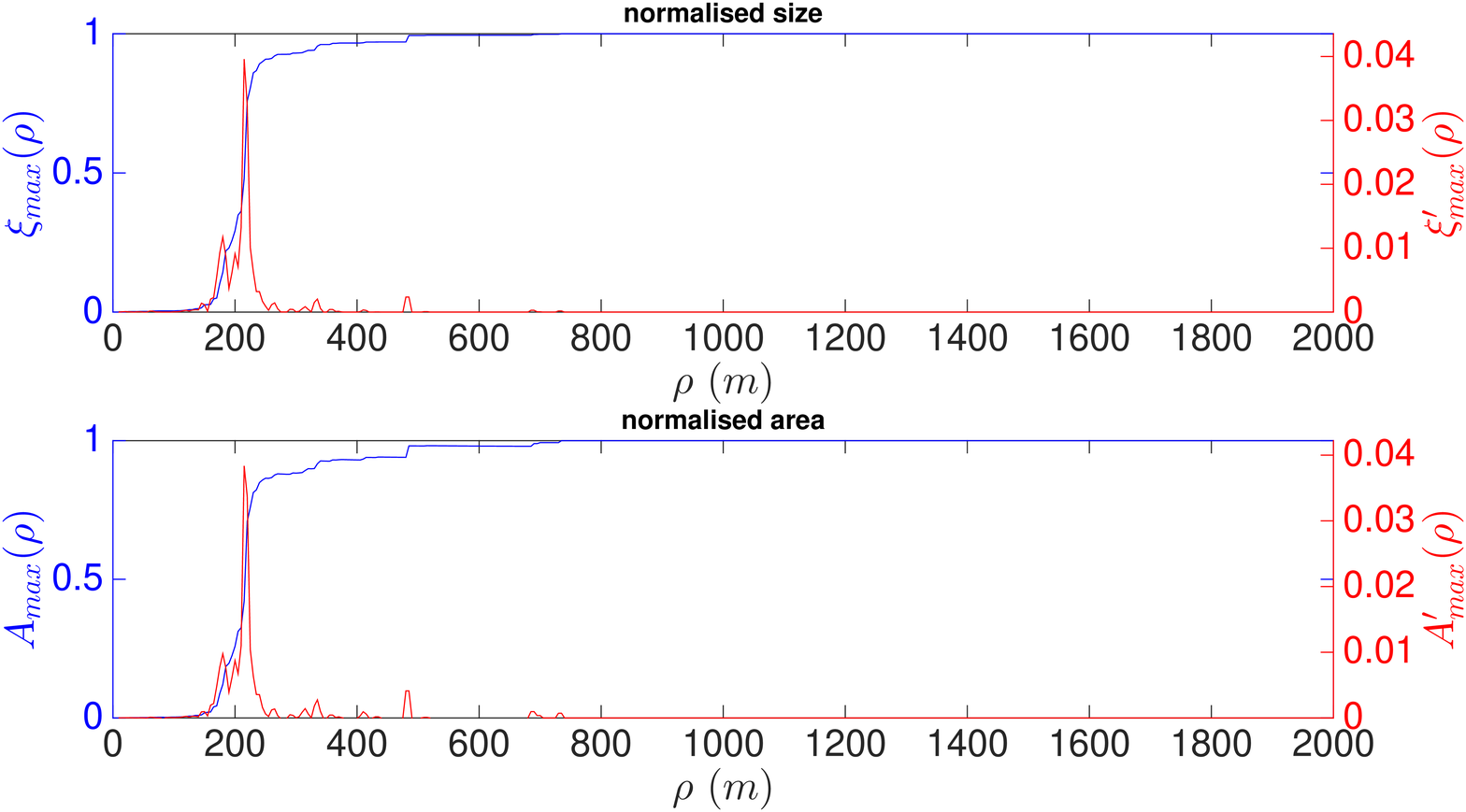}}
\subfigure[\flabel{Epsom}Epsom, Auckland, New Zealand. The spreads of transition
for the size $\sigma_\xi$ and area $\sigma_A$ are not small but stay comparable
to one another. This is an example of multi-scale regularly distributed
pattern.]{%
\includegraphics[width=0.355\textwidth]{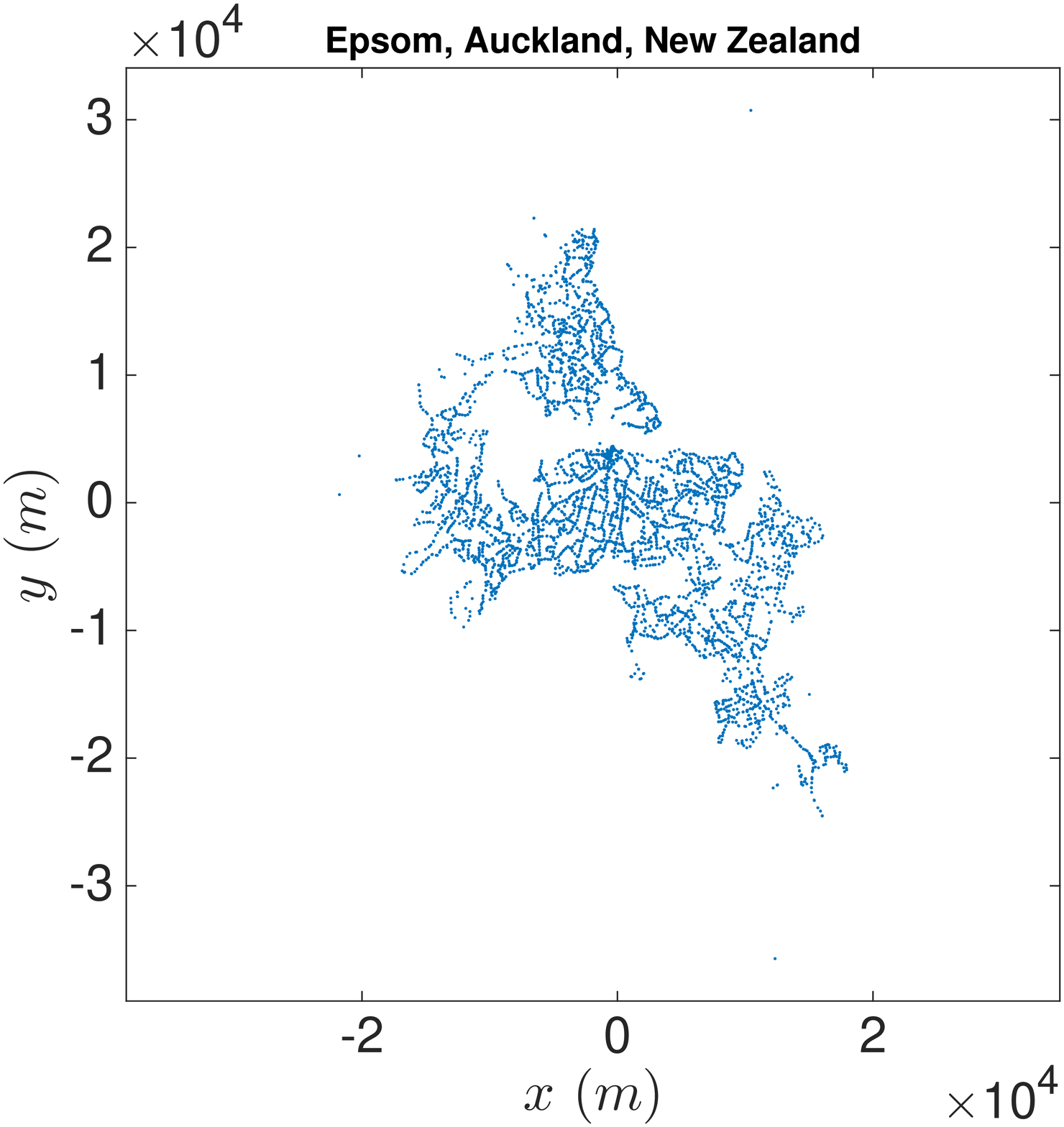}
\includegraphics[width=0.645\textwidth]{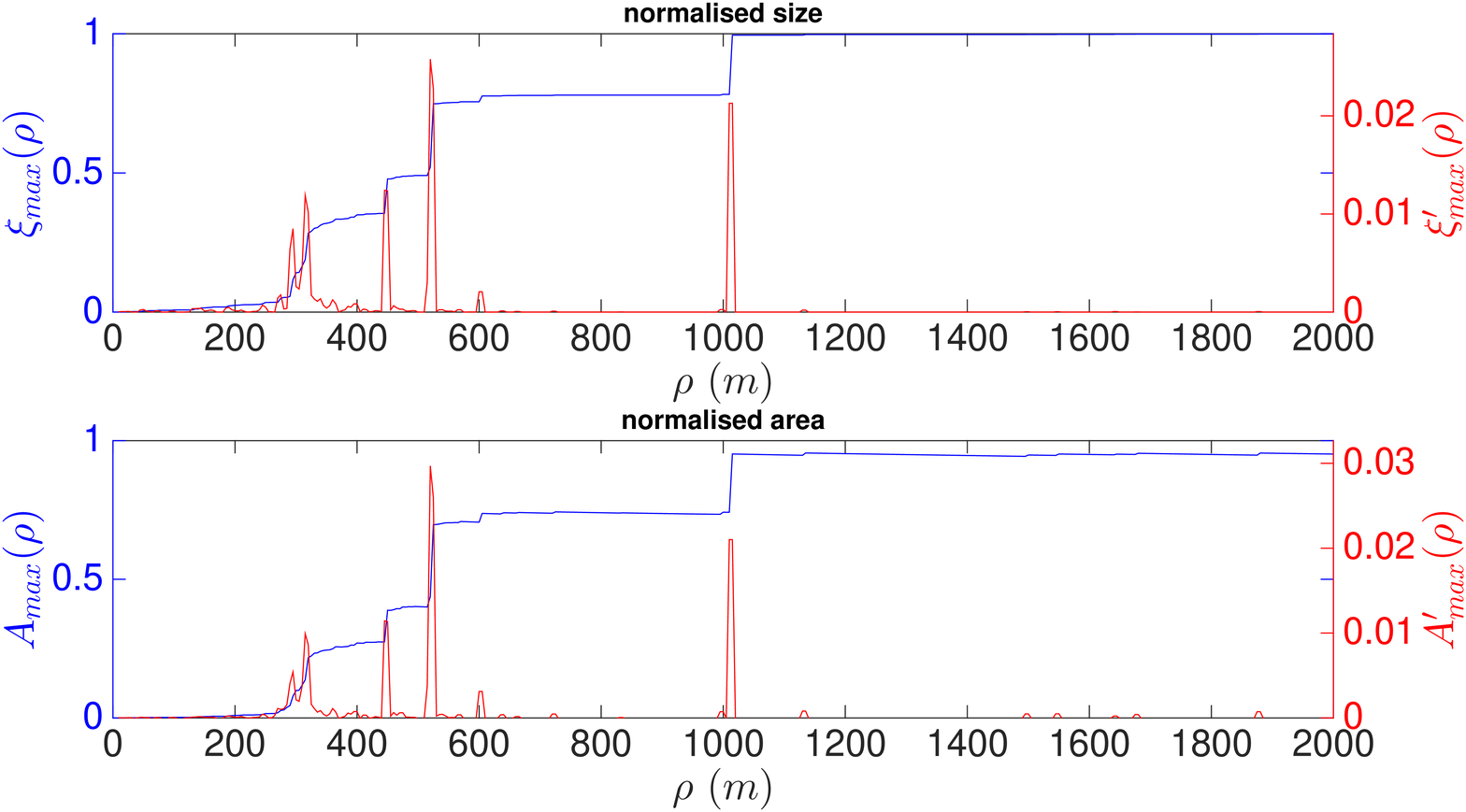}}
\caption{Typical cities of single and multiple-scale regular spatial patterns.
In each subfigure, the left panel shows the location of transport points within
the city, the upper right panel the profile of largest cluster size
$\xi_{max}(\rho)$ together with its first derivative $\xi'_{max}(\rho)$ and the
lower right panel the profile of largest cluster area $A_{max}(\rho)$ together
with its first derivative $A'_{max}(\rho)$}
\end{figure*}

\begin{figure*}[ht!]
\centering
\subfigure[\flabel{Turin}Turin, Piedmont, Italy. The spread of
transition in peaks for the size is more than that for the area of the largest
cluster, $\sigma_\xi>\sigma_A$. This is an example of clustered pattern.]{%
\includegraphics[width=0.355\textwidth]{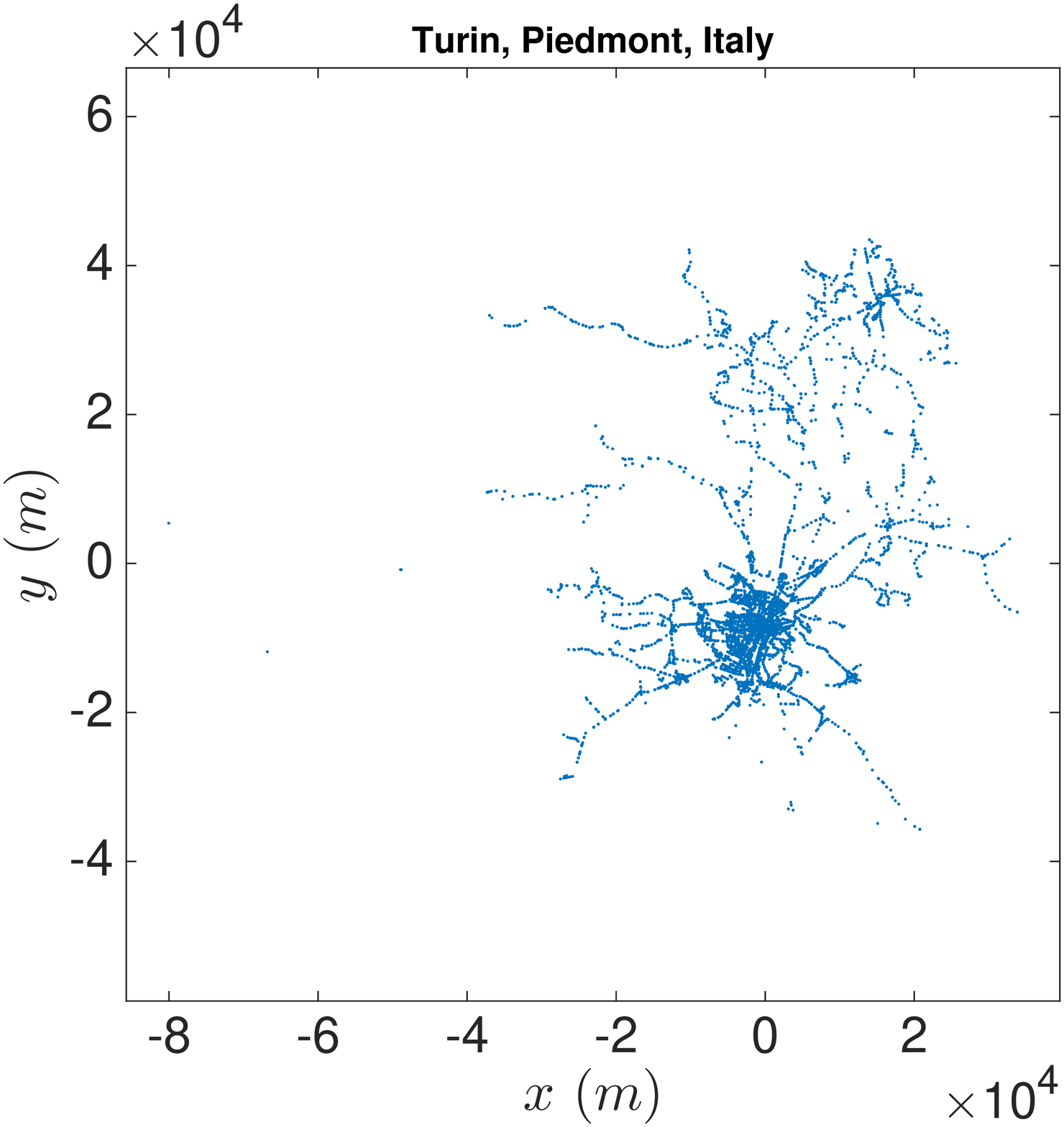}
\includegraphics[width=0.645\textwidth]{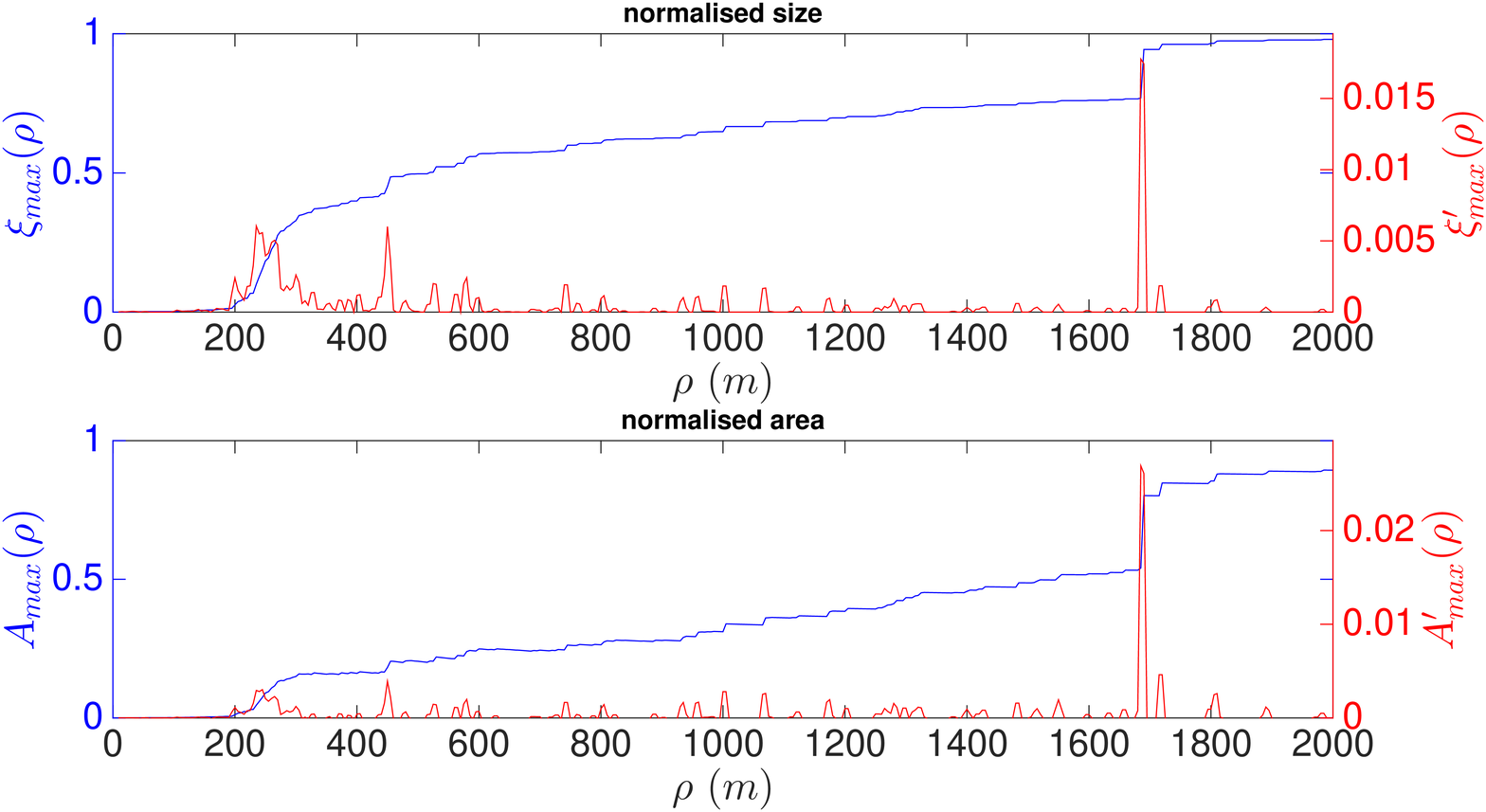}}
\subfigure[\flabel{Manchester}Manchester, Greater Manchester, England. The
spread of transition in the peaks for the size is less than that for the area of
the largest cluster, $\sigma_\xi<\sigma_A$. This is an example of dispersed
pattern.]{%
\includegraphics[width=0.355\textwidth]{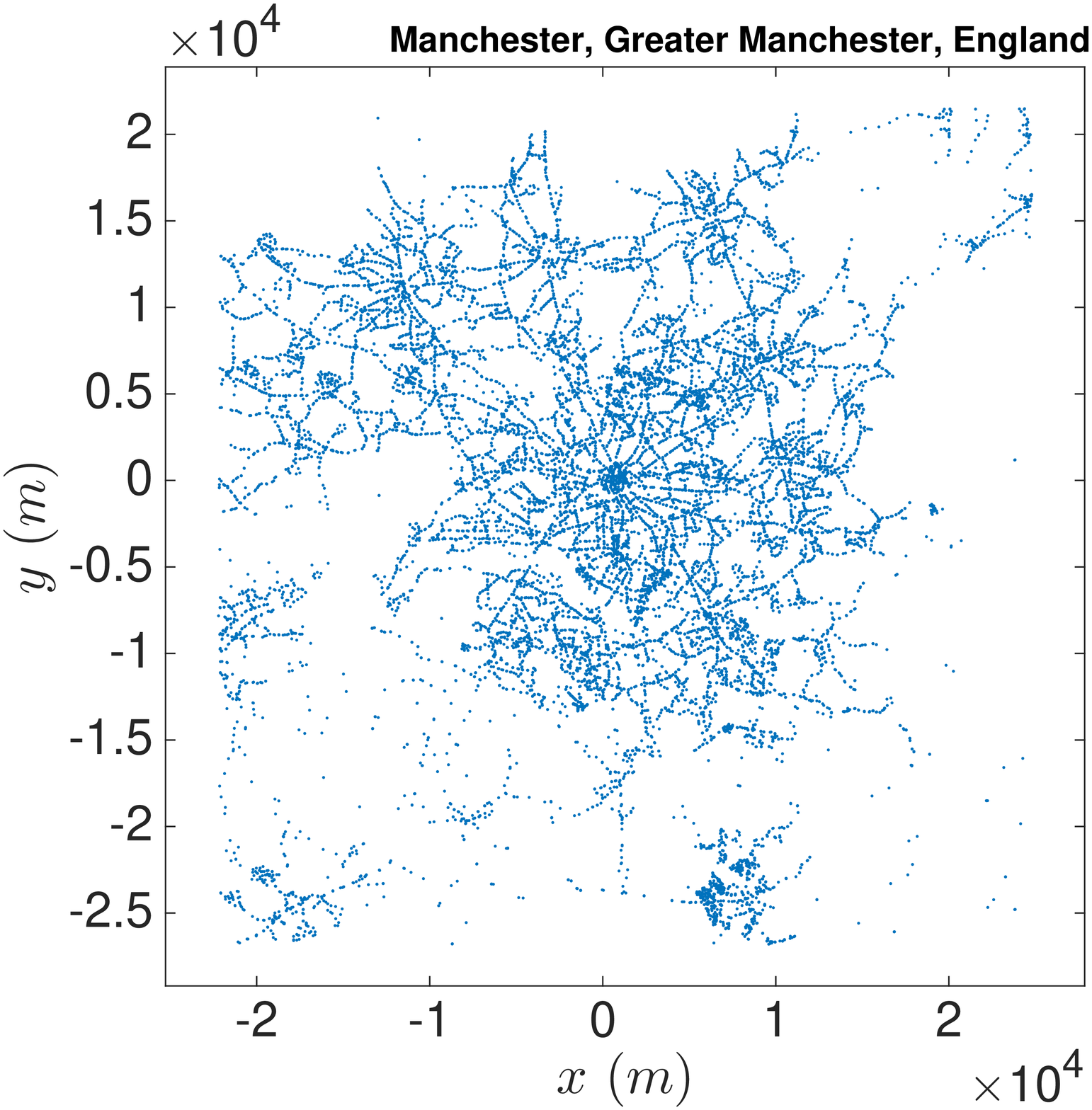}
\includegraphics[width=0.645\textwidth]{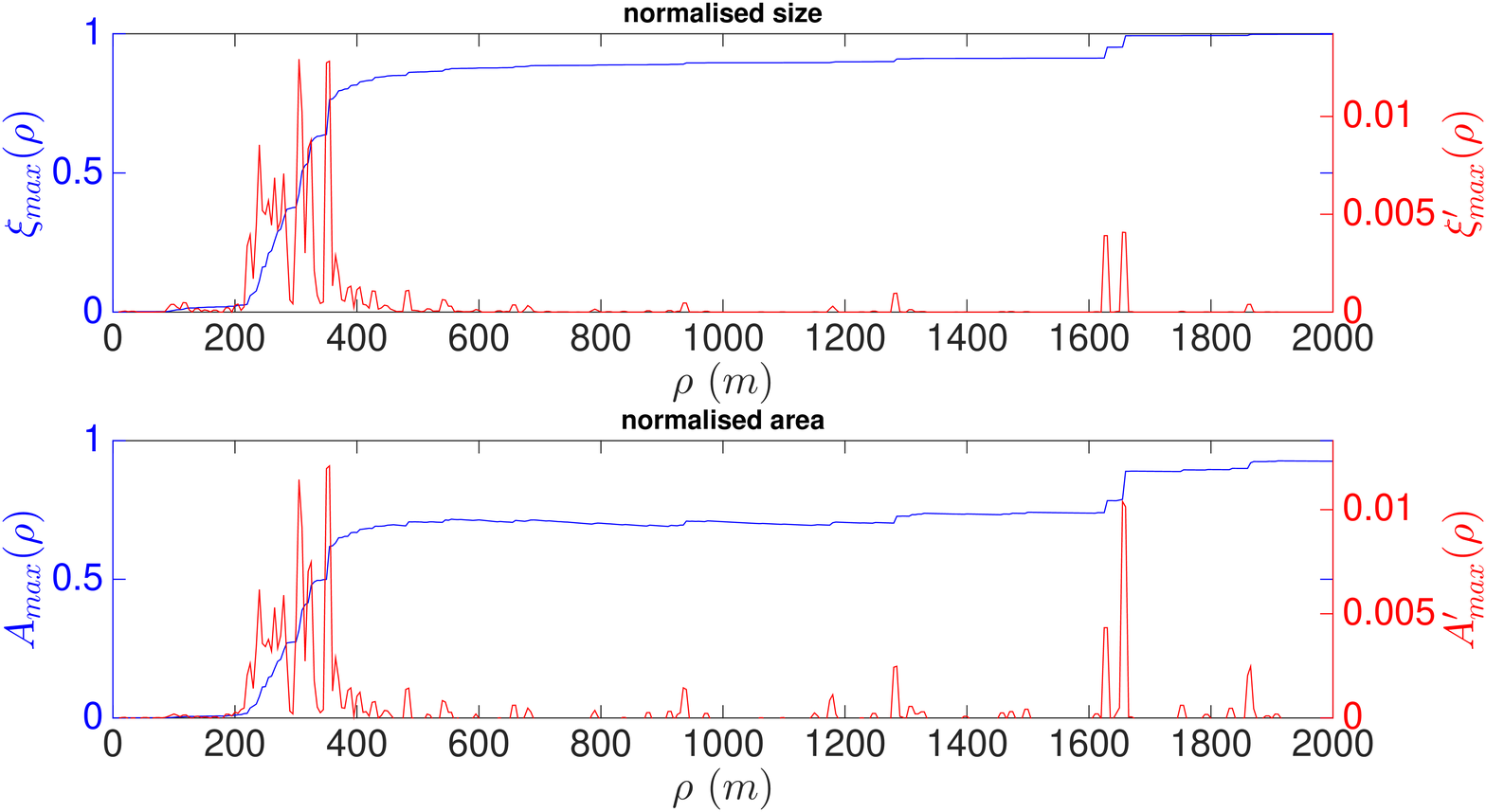}}
\caption{Typical cities of clustered and dispersed spatial patterns. In each
subfigure, the left panel shows the location of transport points within the
city, the upper right panel the profile of largest cluster size
$\xi_{max}(\rho)$ together with its first derivative $\xi'_{max}(\rho)$ and the
lower right panel the profile of largest cluster area $A_{max}(\rho)$ together
with its first derivative $A'_{max}(\rho)$}
\end{figure*}

\subsubsection{Single-scale regular pattern}

In the bottom left corner of the $(\sigma_\xi,\sigma_A)$ plot lie the points
with $\sigma_\xi,\sigma_A<50$. These points represent the profiles of
$\xi_{max}(\rho)$ and $A_{max}(\rho)$ with localised peaks in both
$\xi'_{max}(\rho)$ and $A'_{max}(\rho)$. This signifies a characteristic length
scale at which most of the points are (approximately) equally spaced from each
other, \eg grid points. The boroughs of Bronx, Brooklyn and Manhattan of New
York city are typical examples of such kind of distribution (see
\Fref{Brooklyn}). The spatial pattern of the transport point in these cities
appears very regular. In fact, inspecting their street patterns, one can easily
tell the pattern of parallel roads in one direction cutting those in the other
dividing the land into well organised polygons with almost perfect square and
rectangular shapes. Apparently, this feature must be a result of well-designed
and top-down planning before actually building the infrastructure in the city
\cite{2013@Barthelemy.etal}.

\subsubsection{Multiple-scale regular pattern}

The transport points in a city can also be distributed in a regular manner but
at different length scales. For example, the entire set of points can be divided
into several subsets and within each subset, the points are (quasi-)equally
distant from each others. At larger length scale, \ie $\rho$ increases further,
these subsets of points are again (quasi-)equally distant from each other, \ie
hierarchical structure. The buffer radius $\rho$ can thus be thought to play the
r\^ole of a zooming parameter. In this multi-scale regular pattern, the profile
of the largest cluster size $\xi_{max}(\rho)$ and area $A_{max}(\rho)$
experience a significant jump every time $\rho$ changes its zooming level. At
the lowest level are individual transport point. When $\rho$ zooms out to the
second level, the points that are closest to each other start to form their
respective clusters. Moving to the next level, the nearby clusters start joining
to form larger cluster but there will be many of these ``larger clusters'', \ie
the largest cluster is of comparable size or area to several other clusters. The
most important feature of this spatial pattern is that the jumps in the profile
of $\xi_{max}(\rho)$ correspond well to those in $A_{max}(\rho)$, even though
the locations of the jumps are spread apart. That leads to the (approximate)
equality of the spread of transitions $\sigma_\xi$ and $\sigma_A$ despite their
not being small. A good example of this type of distribution is the city of
Epsom in Auckland, New Zealand (see \Fref{Epsom}).

It is also interesting to note that within a city itself, different parts can
possess distinct spatial patterns of the transport points. For example, even
though New York city is known to be a well-planned city with grid-like street
patterns, not all of its five boroughs share that nice feature. Only Manhattan,
Bronx and Brooklyn have small $\sigma_\xi$ and $\sigma_A$ while the spreads are
larger for the other two boroughs, Queens and Staten Island. This fact indeed
complements the result reported earlier that Queens exhibits a distinct spatial
pattern different from the other boroughs \cite{2014@Louf.Barthelemy}. St Louis
in Missouri, USA, is another interesting example. Two halves of the city on the
two banks of Mississippi river appear to have different spatial patterns when
they possess different values of the pair $\sigma_\xi$ and $\sigma_A$.

\subsubsection{Clustered pattern}

There are cases in which the jumps in the profile of largest cluster size
$\xi_{max}(\rho)$ don't correspond to those in the area $A_{max}(\rho)$ and \vv.
In such cases, the spatial distribution of the transport points deviates from
regular patterns. We first consider the scenarios in which
$\sigma_\xi\gg\sigma_A$. For such distributions, the points are clustered and
tend to minimise the coverage area. When $\sigma_\xi\gg\sigma_A$, there are
jumps in the size of the largest cluster size that do not give rise to a jump in
its area. This happens when the points of an acquired cluster are compact,
contributing very little increase in the area of the largest cluster. If the
acquired cluster are not compact, \ie its points span a larger area, there might
be significant increase in the area of the largest cluster and, hence, a peak
would be reflected by its contribution to $\sigma_A$. However, the size measure
is not affected as it only tells the number of points that are included in the
cluster but not their relative location with respect to each other.

The distribution of transport points in the city of Turin in Piedmont, Italy
(see \Fref{Turin}), is a good example of this type. The points appear clustered
and compactly distributed but not regular or grid-like.

\subsubsection{Dispersed pattern}

On the other side, we have the scenarios of $\sigma_\xi\ll\sigma_A$, in which
the points are dispersed and tend to maximise the coverage area. When
$\sigma_\xi\ll\sigma_A$, there are jumps in the area of the largest cluster that
do not give rise to a jump in its size. This happens when the points of an
acquired cluster are dispersed (but still within the buffer radius so that they
belong to the same cluster). This way, the increase in the area of the largest
cluster is more significant than that in its size, resulting
$\sigma_\xi\ll\sigma_A$. A good example of this type is the distribution of
transport points in Manchester in Greater Manchester, England (see
\Fref{Manchester}). The points appear in dispersed pattern of long roads around
the city.

If the feature of single-scale regular spatial pattern (when both $\sigma_\xi$
and $\sigma_A$ are small) is a result of well-designed and top-down planning in
an urban system, the other spatial patterns (either $\sigma_\xi$ or $\sigma_A$
is not small) can be intepreted as a consequence of developing an urban system
under local constraints. In the former case, the urban system appears to be of
organised type while in the latter, it can be said to be of organic type when
its spatial features develop in an adhoc manner as the city grows. The
revelation of spatial patterns in urban systems from the analysis in this work
could imply two different types of process that the cities undergo through their
course of development.

Visually inspecting the spatial distribution of transport points within the
cities, it appears that cities with regularly distributed pattern, either
single- or multiple-scale, do not have an apparent centre. That means there is
no spatial preference in the distribution of the points, \ie no part is special
than the others. This is in contrast to the other two types of cities in which
star-node structure can be clearly observed. The node represents the centre of
the city at which there is higher density of transport points than the other
areas, and from which the roads diverge radially to the outer part of the city.
This observation could by explained by the growth process of different types of
urban system. When a city grows organically, it starts from a central business
district and gradually expands to encompass the nearby area to accommodate more
people wanting to participate the business activities at the centre. On the
other hand, when a city is planned before, the planners seem not to concentrate
the infrastructure in one confined area but stretch it across the entire city.

\subsection{Universal features in urban systems}
\Slabel{universal_features}

\subsubsection{Area of system of transport points}
\Slabel{density_distance_relation}

Despite the difference in the spatial pattern of the distribution of transport
points, all the cities considered in this study appear to possess a common
relation between the density of points and the characteristic distance of area
$\rho_A^\star$. To explore this, we first define the characteristic area of a
set of points which is the union area of circles of radius $\rho_A^\star$
centred at all points in the set, $\Xi(\rho_A^\star)$. The characteristic
distance of area $\rho_A^\star$ represents the typical value of $\rho$ at which
the area of the entire system experiences the transition. Further increase in
$\rho$ after that does not contribute to as much increase in the area of the
largest cluster and the cluster, thus, would entail unnecessary area. The value
of $\Xi(\rho_A^\star)$ is therefore expected to be a good proxy to the essential
area covered by the set of points given their spatial distribution. In the case
of transport system, we call this area the serving area of all the transport
points.

Having defined the area, it is easy to calculate the density of points per unit
area, which is simply the ratio between the number of points and the area they
cover, $\displaystyle\frac{N}{\Xi(\rho_A^\star)}$. From empirical analysis, we
find the data fit very well to the relation
\begin{equation}
\elabel{density_relation}
\frac{N}{\Xi(\rho_A^\star)} \propto \left(\rho_A^\star\right)^{-\tau}
\end{equation}
with $\tau\approx1.29$. \BFref{density_relation} shows the empirical relation
between the two quantities. The relation in \Eref{density_relation} is well
obeyed by all $73$ cities. It is a remarkable relation given the scattered
relation between $\rho_A^\star$ and $N$ or $\Xi(\rho_A^\star)$. For reference,
the artificial generated data, including both random and regular patterns (see
\Sref{discussion} for details) are also included in the plot but not in the
fitting itself. As can be observed, those points generally stay below the points
for the $73$ cities.

\begin{figure*}[ht!]
\centering
\includegraphics[width=1\textwidth]{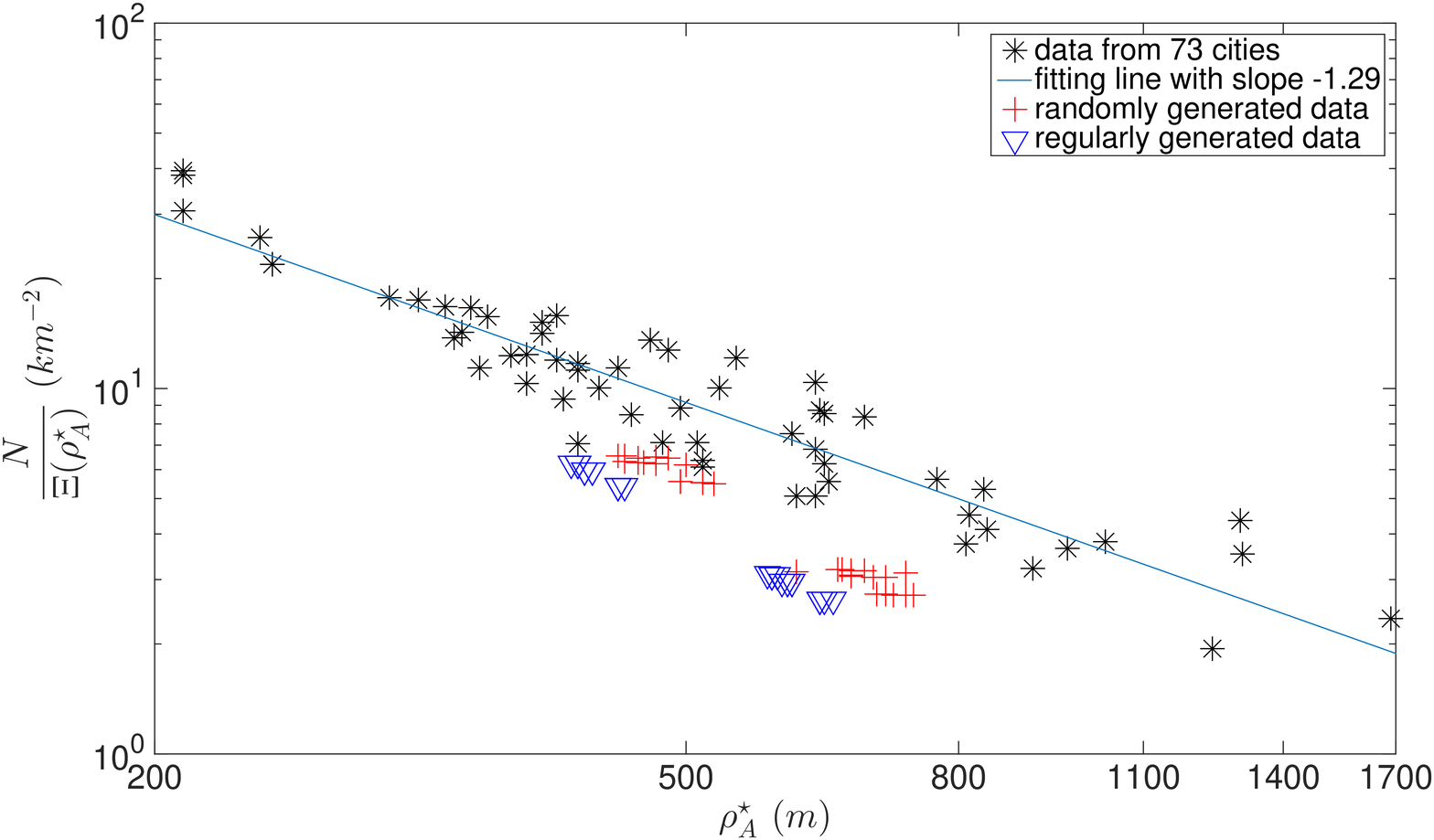}
\caption{\flabel{density_relation}Empirical relation between density of point
per unit area and the characteristic distance $\rho_A^\star$ in log-log scale.
The fitting line has slope $-1.29$ and was obtained with linear regression
coefficient of $R^2\approx0.8$. The artificial generated data, both random and
regular, were not included in the fitting.}
\end{figure*}

It should be further noted that the relation in \Eref{density_relation} or
\Fref{density_relation} is not a trivial one. To see this, we consider the two
scenarios of $\rho$ in extreme limits, $\rho\gg1$ and $\rho\ll1$, and its
relation with $\displaystyle\frac{N}{\Xi(\rho)}$. For the small extreme value,
$\rho\ll1$, all clusters include only a single point, and there are, hence, $N$
clusters. We, therefore, have
\begin{equation}
\Xi(\rho) = N\pi\rho^2
\end{equation}
which yields the relation
\begin{equation}
\frac{N}{\Xi(\rho)} = \frac{1}{\pi\rho^2} \propto \rho^{-2}\text{.}
\end{equation}

At the other extreme value, $\rho\gg1$, there is only one single cluster that
encompasses all points concentrating at the centre of the union area. We,
therefore, have
\begin{equation}
\Xi(\rho) \approx \pi\rho^2\text{,}
\end{equation}
which leads to
\begin{equation}
\frac{N}{\Xi(\rho)} \approx \frac{N}{\pi\rho^2}\text{,}
\end{equation}
which in turn displays scaling behaviour like in \Eref{density_relation} if and
only if the number of points $N$ scales with $\rho$.

The whole argument about the extreme values of $\rho$ is to illustrate that the
scaling relation in \Eref{density_relation} with exponent $\tau=-1.29$ is not a
relation that can be achieved with any value of $\rho$. The relation can only
hold at some value of the buffer radius like $\rho_A^\star$, given the
structure in the distribution of $N$ points. Because of this feature, we
consider $\rho_A^\star$ \emph{the} characteristic distance of a set of
spatially distributed points.

\subsubsection{Amenity distribution}

Beside transport system, which is represented by a network of transport points,
the morphology of an urban system can also be understood from another angle by
examing the distribution of amenities within it. It turns out that despite
possessing different types of distribution of transport points, the cities
appear to share a common universal distribution of amenities. The analysis of
locations of amenities in all the cities reveals that the (Euclidean) distance
$\Omega_k$ of an amenity $k$ to its nearest transport point follows a robust
exponential distribution. That means the probability of finding an amenity with
distance $\Omega$ to its nearest transport point decays exponentially with
$\Omega$, \ie its probability density function is given by
\begin{equation}
P(\Omega) = \lambda\operatorname{e}^{-\lambda\Omega}\text{,}
\end{equation}
which renders its mean and standard deviation (\emph{not variance}) equal
\begin{equation}
\ave{\Omega} = \sigma(\Omega) = \sqrt{\ave{\Omega^2}-\ave{\Omega}^2} =
\frac{1}{\lambda}\text{.}
\end{equation}

In \Fref{mean_stddev}, the mean $\ave{\Omega}$ and standard deviation
$\sigma(\Omega)$ of shortest amenity-transport point distance for different
cities are shown to stay close to the diagonal line
$\sigma(\Omega)=\ave{\Omega}$. The exponential distribution of the distance
$\Omega$ is strongly supported by further veriyfing that higher moments of the
distribution $P(\Omega)$ fit well to
\begin{equation}
\ave{\Omega^n} = \frac{n!}{\lambda^n}\text{,}
\end{equation}
up to fourth order, $n=4$.

\begin{figure*}[ht!]
\centering
\includegraphics[width=\textwidth]{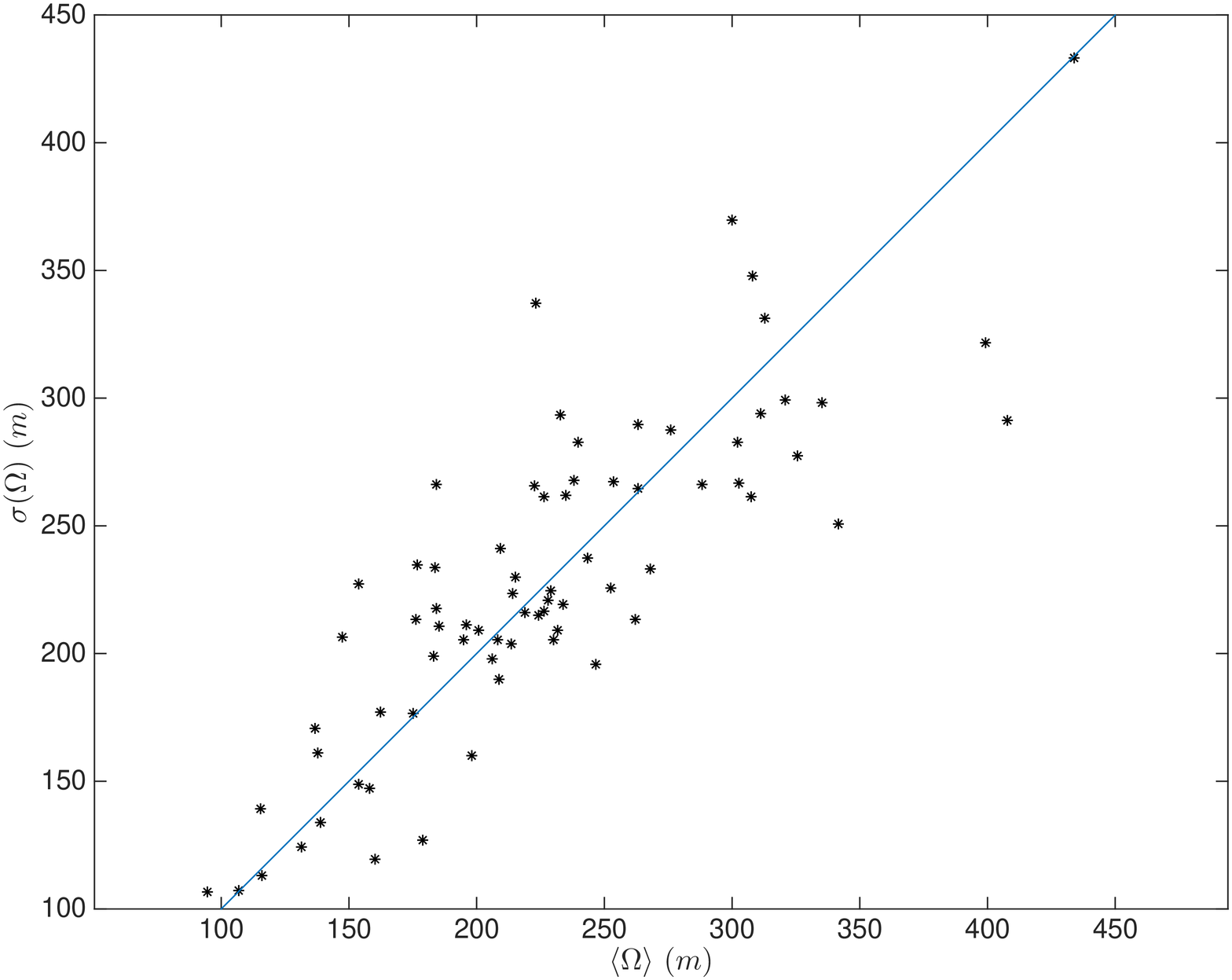}
\caption{\flabel{mean_stddev}Standard deviation $\sigma(\Omega)$ \vs mean
$\ave{\Omega}$ of distance from amenities to nearest transport points within
each city. The reference line is $\sigma(\Omega)=\ave{\Omega}$.}
\end{figure*}

It has to be emphasized that the distribution of distance from amenities to
their nearest transport points follows an exponential rather than a Poissonian
one. That means the mean of such distance is (approximately) equal to its
standard deviation rather than variance which holds for a Poisson distribution.
The robust distribution of amenities across all city types has one important
implication that the local growth process in urban systems appears to be
independent of human intervention and larger scale of the entire system. That
means planners can plan the large-scale growth process like transportation but
the small-scale growth process like local business still takes place on its own.
But it remains a significant question why the distribution is exponential, not
any other form. In fact, exponential decay in spatial urban patterns has been
long reported in literature \cite{1951@Clark}. Using this feature as a fact, a
model has been constructed to successfully capture the morphology of urban
systems \cite{1998@Makse.etal}.

\subsection{Relation between transport point and amenity distributions}

There appears to be a relation between the density of transport points within a
city and the distribution of its amenities. \BFref{density_distance} depicts
this relation by plotting the average distance $\ave{\Omega}$ agaisnt the
density $\displaystyle\frac{N}{\Xi(\rho_A^\star)}$. It could be observed that
only the lower-left triangle of the plot is occupied, leaving no points in the
upper-right corner of the plot. That means there are no cities with high density
of transport points and, at the same time, having large (average) distance
between its amenities and the nearest transport points. This can be easily
understood by the fact that in cities with high density of the transport points,
the road network is very dense, the transport points have to stay within a short
distance of each other. As a result, the amenities must necessarily be built
very close to the public transport points. It turns out that those cities with
very high density of transport points are those with single-scale regular
pattern of distribution, \ie the points are regularly distributed at
(approximately) equal distances from each other like grid points, such as San
Francisco or the boroughs of New York city.

\begin{figure*}[ht!]
\centering
\includegraphics[width=\textwidth]{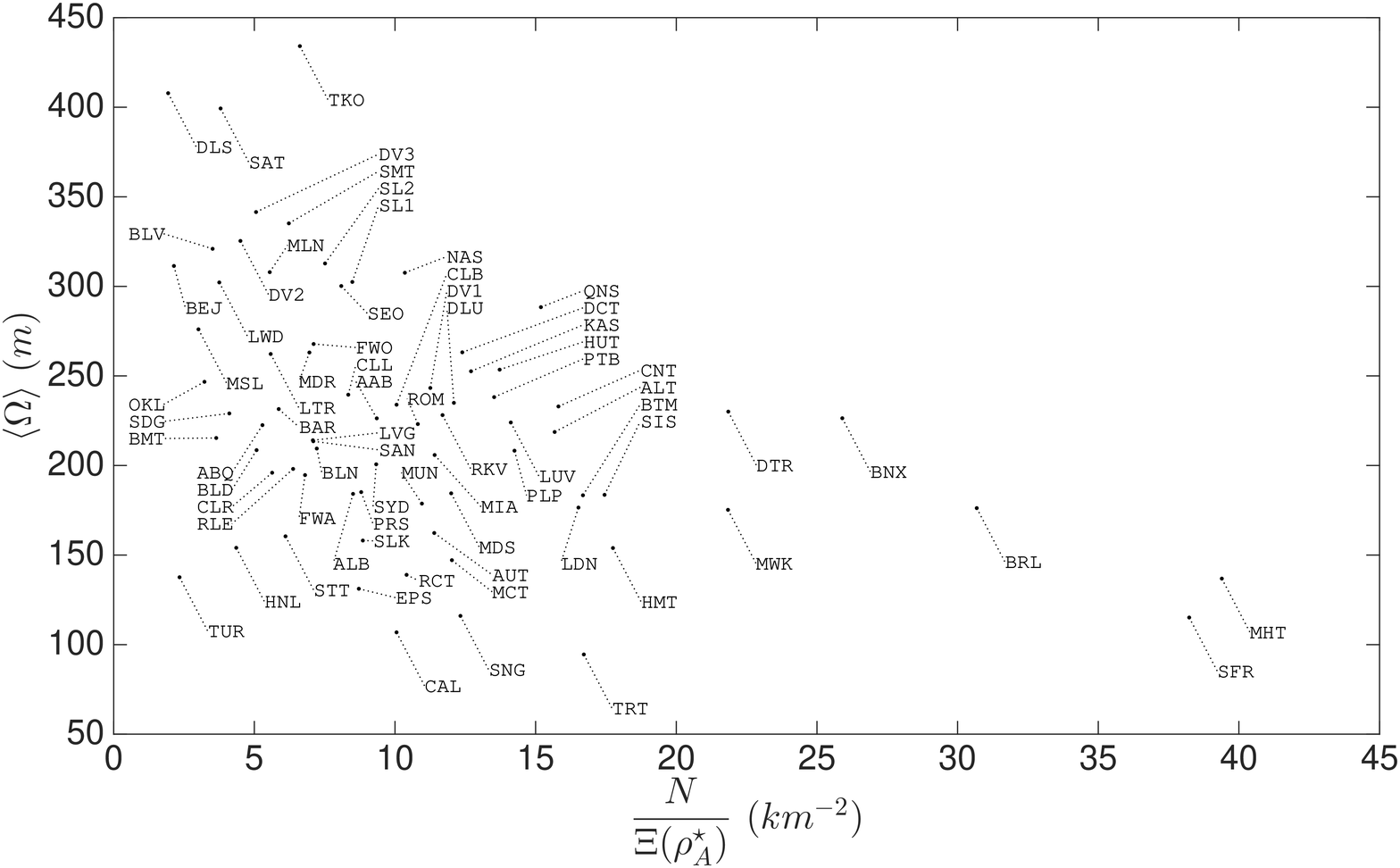}
\caption{\flabel{density_distance}Relation between density of transport points
$\displaystyle\frac{N}{\Xi(\rho_A^\star)}$ and average amenity-transport point
distance $\ave{\Omega}$. Refer to \Fref{ss_sa} for codename of the cities.}
\end{figure*}

At the other end, the cities with low density of public transport points can
exhibit a wide spectrum of average amenity-transport point distance
$\ave{\Omega}$. These cities can either have large or small $\ave{\Omega}$. A
large value of $\ave{\Omega}$ (and hence, large standard deviation
$\sigma(\Omega)$, too) implies a city with sparse distribution of amenities when
they are distant from the nearest transport point like Dallas or San Antonio in
Texas, USA. On the other hand, a small value of $\ave{\Omega}$ suggests that the
amenities are built close to public transport points implying the existence of
sub-centres or several small towns or districts within the city, such as Turin
in Piedmont, Italy.

\section{Discussion}
\Slabel{discussion}

The present work analyses the features of the spatial distribution patterns of
important entities in an urban system, the public transport points and the
amenities. The former ones are part of the backbone of any urban system, the
street network which plays essential r\^ole in enabling flow or exchange of
various processes in the city. The advantage of knowledge of these transport
points is that they can be well defined and easily collected and at the same
time provide other information like the residential distribution within the
city. The latter ones, on the other hands, can gauge the size of population as
well as the level of activities in the city. The results unveil different types
of city with distinct spatial patterns. The cities are shown to be either of
organised type, in which the entities are well spaced as if they are built
top-down, or of organic type, in which the entities are spaced with multiple
length scales as if they grow spontaneously. In either cases, the typical
distance among the transport points can be described by a characteristic
distance. Despite the different types of the cities' spatial patterns, the
density of the transport points exhibits universal scaling behaviour with this
characteristic distance. On the other hand, the distance from amenities to their
nearest transport points also follows a robust exponential distribution for all
cities studied. Furthermore, there is an apparent relation between the
distributions of transport points and amenities within the cities. These facts
signify some universal mechanisms underlying the growth and development that all
cities have to undergo.

In an attempt to understand the processes that generate the spatial distribution
patterns of the transport points, we artificially generate some distributions of
points on a two-dimensional surface. In the first distribution, the points are
generated at random locations within a domain with uniform probability. In the
second distribution, starting from a regular grid of points in a square lattice,
the points are randomly displaced by a small amount not more than a quarter of
the lattice spacing. It turns out that both the randomly and regularly generated
data produce simple behaviours through our analysis. In particular, the peaks in
size and area generally coincide with each other and stay localised (the random
sets tend to produce more peaks while the regular ones have only one peak as
expected), and hence, both $\sigma_\xi$ and $\sigma_A$ are small indicating
regular pattern of distribution. At this point, we would like to link the
analysis with the idea of measure of complexity of a symbolic sequence
\cite{1989@Crutchfield.Young,2012@Crutchfield,2015@Huynh.etal}. The idea states
that both regular and random (in the sense of a random number generator)
sequences possess very low measure of complexity as their structures or patterns
are simple and easy to be presented in terms of the so-called $\epsilon$-machine
\cite{1989@Crutchfield.Young}. Along that line, it could be argued that the
patterns observed in the distributions of transport points from the real data of
$73$ cities around the world are more complex then those in the artificially
generated data. The generated data, which is meant to be either regular or
completely random possess, only simple structure as we have argued above with
small spreads $\sigma_\xi$ and $\sigma_A$. The real world data might well
contain mixtures between regular and random patterns that could result in both
the clustered and dispersed patterns that we have reported in
\Sref{spatial_patterns}.

\footnotesize

\section{Method}
\subsection{General ideas}
For the analysis, we propose a procedure to characterise the spatial pattern of
a set of points. The procedure involves identifying clusters of points, whose
pairwise distance does not exceed the value of a parameter, and quantifying the
growth of the clusters as the parameter value increases.

Consider a domain $\mathfrak{D}$ which can be thought of as a city or a town. In
this domain, there are $N$ points $i$ distributed, each of which represents a
transport point located at coordinates $(x_i,y_i)$. We introduce a parameter
called the buffer radius $\rho$ to construct the clusters. Any point $j$, whose
distance
\begin{equation}
d_{ij} = \sqrt{(x_i-x_j)^2+(y_i-y_j)^2}
\end{equation}
from point $i$ is less than or equal to $\rho$, belongs to the same cluster as
$i$. We denote $\eta(\rho)$ as the number of clusters given the buffer radius
$\rho$. For each cluster $\alpha$, we define the cluster size $\xi_\alpha(\rho)$
and the cluster area $A_\alpha(\rho)$. The cluster size is defined as the number
of points in the cluster and the cluster area the union of area of circles of
uniform radius $\rho$ centred at the points in the cluster. To make different
domains comparable, we normalise the cluster size $\xi_\alpha(\rho)$ by the
number of points $N$ in the domain, while the cluster area $A_\alpha(\rho)$ by
the union of area $\Xi(\rho)$ of circles of radius $\rho$ centred at all points
in the domain.

The identification of the clusters can be done by using a simple heuristic
cluster finding algorithm that starts with a random points in the set and
gradually identifies the other points of in the same cluster. Alternatively, one
can employ the method of DBSCAN \cite{1996@Ester.etal}, setting the noise
parameter to be zero. The two methods are identical and yield the same results.

\subsection{Analysis}
For any cluster-related quantity, we attach the subscript $\xi$ to associate it
with cluster size while $A$ for cluster area. For simplicity of all discussions,
unless stated explicitly, descriptions for cluster size $\xi$ also hold for
cluster area $A$.

To quantify the spatial pattern of the set of points, we vary the buffer radius
parameter $\rho$. As $\rho$ increases, the farther points can belong to the same
cluster. As a result, the clusters can merge to increase their size. Tracing the
behaviour of the largest cluster $\xi_{max}(\rho)$ can provide us with the way
the points are distributed within the set. For example, the profile of the first
derivative
$\displaystyle\xi'_{max}(\rho)=\frac{\dint\xi_{max}(\rho)}{\dint\rho}$ (and
$\displaystyle A'_{max}(\rho)=\frac{\dint A_{max}(\rho)}{\dint\rho}$) can
indicate at which distance $\rho$, the points are (largely) connected in a
single cluster. Because $\xi_{max}(\rho)$ increases monotonically with $\rho$,
we introduce the so-called \emph{characteristic distance} $\rho_\xi^\star$ at
which $\xi_{max}(\rho)$ exhibits the most significant increase. In some cases,
the profile $\xi_{max}(\rho)$ shows a sharp narrow increase around a value
$\rho$. While in other cases, several small increases are observed, spreading a
wide range of $\rho$. To account for that, it is also meaningful to introduce a
quantity $\sigma_\xi$, called \emph{spread of transition}, to measure the
overall width of the increases in the profile of $\xi_{max}$.

To recap, $\rho_\xi^\star$ is the value of $\rho$ above which there is a
significant transition in the largest cluster size $\xi_{max}(\rho)$ (see peak
analysis in \Sref{peak_analysis}); $\sigma_\xi$ measures the width of the
transition. The respective quantities for cluster area are $\rho_A^\star$ and
$\sigma_A$. We then use the union area with this buffer radius $\rho_A^\star$ of
all points in the system as the effective area $\Xi(\rho_A^\star)$ of the point
set.

We refrain from looking at the average value of distribution of the cluster size
or area as these measures are vulnerable against errors in data, \ie outliers.
For example, when the buffer radius $\rho$ is sufficiently large that most but a
few of the points in the dataset belong to a single large (``giant'') cluster,
there are only two (or more) clusters and the average cluster size would be half
(or less) of what it is supposed to be.

\subsubsection{Peak analysis}
\Slabel{peak_analysis}
As the buffer radius $\rho$ increases, the size of the largest cluster
$\xi_{max}(\rho)$ also increases. It can be easily observed that the profile of
$\xi_{max}(\rho)$ exhibits an either sharp or gradual increase. The former
introduces a single dominant peaks in the profile of $\xi'_{max}(\rho)$ while
the latter a set of peaks scattering over a wide range of $\rho$. This
scattering of peaks can be quantified using the standard deviation of their
locations, weighted by the strength (height) of the peaks. A small standard
deviation implies a sharp increase in $\xi_{max}(\rho)$, and \vv, a large
standard deviation signifies a gradual increase.

In our analysis, we consider the profile of $\xi'_{max}(\rho)$ (and
$A'_{max}(\rho)$) at every value of the buffer radius $\rho$ ranging from
$\rho_{min}=\rho_1=10\text{m}$ to $\rho_{max}=\rho_M=2000\text{m}$ in the step
of $\delta\rho=\rho_{i+1}-\rho_i=5\text{m},~\forall i$. Since the values of the
buffer radius are discrete, a point $(\rho_i,\xi'_{max}(\rho_i))$ is a peak if
and only if
\begin{equation}
\left\{
\begin{aligned}
\xi'_{max}(\rho_i) & >\xi'_{max}(\rho_{i-1}) \\
\xi'_{max}(\rho_i) & >\xi'_{max}(\rho_{i+1})
\end{aligned}
\right.\text{.}
\end{equation}
This discrete nature also produces a lot of small noisy peaks. In our analysis,
we filter these noisy peaks by offsetting the entire profile of
$\xi'_{max}(\rho)$ by a sufficiently small amount and considering only the
positive remaining peaks.

The value of $\rho^\star_\xi$ will then be the mean of $\rho$ of all peaks,
weighted by the peak height $\xi'_{max}(\rho)$ (see
\Eref{characteristic_distance}). The spread $\sigma_\xi$ is the standard
deviation of $\rho$ of all peaks, again weighted by the peak height
$\xi'_{max}(\rho)$ (see \Eref{spread_of_transition}).

\subsection{Implementation}

The cluster analysis is implemented in Python, using \texttt{shapely} library to
calculate the union area. We also use \texttt{DBSCAN} library for DBSCAN
analysis to compare with our method.

\section{Data}
\Slabel{data}

\subsection{Data collection}

The public transit-related data was collected by Baseride Technologies using
different available API (application programming interfaces) provided by transit
agencies. Collected data included (but not limited to) GPS (global positional
system) coordinates of bus stops, their characteristics (\eg name), route
geometry, bus stop sequence on the routes. Additional information was also
collected for future analysis (schedules, trips, real time public transport
location updates). Majority of agencies provides information through GTFS
(general transit feed specification). Minority of cities are using their
custom-made APIs. Baseride converted all protocols into single uniform data
representation. Information about public transit network was also converted into
linked graph for convenient analysis using different methods. The data for
Singapore was obtained from Transit Link Pte Ltd \cite{transitlink}. The data
set was also augmented with data from OSM.

The amenity data for all cities was obtained from OSM through the Mapzen project
\cite{mapzen}. An amenity is said not to belong the city if it is not within
$1,000$m of any bus stop in that city.

The datasets contain information about location of the bus stops in the form of
latitude and longitude. The stops are grouped for the same city or municipal
organisation. In the present analysis, we only select $73$ cities with at least
$1,000$ bus stops. Those include the cities in England, France, Germany, Italy,
Spain, Canada, United States, China, Japan, South Korea, Australia, New Zealand
and Singapore. The full list of cities can be found in \Fref{ss_sa}.

\subsection{Data preprocessing}

For each dataset, the spherical coordinates of each bus stop in latitude
$\theta$ and longitude $\varphi$ are transformed to quasi-planar two-dimensional
Cartesian coordinates. We could have done the transformation by employing the
Universal Transverse Mercator (UTM) conformal projection, but since all datasets
are confined within areas on Earth's surface spanning less than $100$km in both
dimensions, we find the approximation method below sufficient, with errors being
less than $0.5\%$\cite{latlong_accuracy}.

We convert the spherical coordinates $\bm\phi_i=(\varphi_i,\theta_i)$ to
Cartesian coordinates $\bm r_i=(x_i,y_i)$ for every point $i$ in the dataset by
first setting the origin of the plot. The origin $O$ is the centroid of all
points
\begin{equation}
\bm\phi_O = \ave{\bm\phi_i} = \frac{1}{N}\sum_{i=1}^N{\bm\phi_i} \text{,}
\end{equation}
for which $\bm r_O=(0,0)$. The Cartesian coordiantes of a point $i$ is then
determined based on its great-circle distance from the origin $O$. In
particular, the $x$-coordinate of $i$ is its (signed) great-circle distance from
the point that has the same longitude $\varphi_O$ as $O$ and the same latitude
$\theta_i$ as $i$ itself. On the other hand, the $y$-coordinate of $i$ is its
(signed) great-circle distance from the point that has the same latitude
$\theta_O$ as $O$ and the same longitude $\varphi_i$ as $i$ itself. The
great-circle distance is calculated using the ``haversine'' formula and, hence,
the Cartesian coordinates of point $i$ are given by
\begin{align}
x_i & =2R\tan^{-1}{\left(
\frac{\cos{\theta_i}\sin{\displaystyle\frac{\varphi_i-\varphi_O}{2}}}
     {\sqrt{1-\cos^2{\theta_i}
              \sin^2{\displaystyle\frac{\varphi_i-\varphi_O}{2}}}}
\right)}\text{,} \\
y_i & =R(\theta_i-\theta_O)\text{,} \\
\end{align}
with $R=6,371,000$m being the Earth's radius.

\normalsize

\bibliography{references}

\section*{Acknowledgements}

We thank Muhamad Azfar Bin Ramli for his help in collecting Singapore bus stop
data.

\section*{Author contributions statement}

HNH, CM and LYC conceived and designed the study. HNH devised the method of
analysis and analysed the data. EM collected the bus stop data of all cities.
HNH and EFL collected the remaining data. HNH wrote the manuscript. All authors
reviewed the manuscript.

\section*{Additional information}

Competing financial interests: The authors declare no competing financial
interests.

\end{document}